\documentclass[review]{elsarticle}

\usepackage{lineno,hyperref,amsmath,amssymb}
\usepackage{lipsum}
\modulolinenumbers[5]

\journal{Advances in High Energy Physics}

\begin{document}

\begin{frontmatter}

\title{Bremsstrahlung from relativistic heavy ions in a fixed target experiment at the LHC}

\author{Rune E. Mikkelsen}
\author{Allan H. S{\o}rensen}
\author{Ulrik I. Uggerh{\o}j}
\address{Department of Physics and Astronomy, Aarhus University\\ Ny Munkegade 120, 8000 Aarhus C, Denmark}

\begin{abstract}
We calculate the emission of bremsstrahlung from lead and argon ions in A Fixed Target ExpeRiment (AFTER) that uses the LHC beams.
With nuclear charges of $Ze$ equal $208$ and $18$ respectively, these ions are accelerated to energies of $7$ TeV$\times Z $.
The bremsstrahlung peaks around $\approx 100$ GeV and the spectrum exposes the nuclear structure of the incoming ion. 
The peak structure is significantly different from the flat power spectrum pertaining to a point charge. 
Photons are predominantly emitted within an angle of $1/\gamma$ to the direction of ion propagation.
Our calculations are based on the Weizs\"{a}cker-Williams method of virtual quanta with application of existing experimental data on photonuclear interactions.

\end{abstract}

\begin{keyword}
\texttt{Bremsstrahlung}\sep \texttt{Ultra-relativistic ions} \sep \texttt{Fixed-targets} \sep \texttt{Ultra-high energy}
\end{keyword}

\end{frontmatter}


\section{Introduction}
The structure of stable nuclei, in particular the charge distribution, may be investigated by impact of photons and electrons as e.g. shown in pioneering works by Hofstadter and collaborators, see e.g.\cite{Hofs56}. This method, however, is not possible for unstable nuclei with short lifetimes as e.g. hypernuclei. Instead, essentially with a change of reference-frame, one may let the nucleus under investigation impinge on a suitable target, e.g. an amorphous foil, and measure the delta-electrons and/or photons emitted in the process. The interaction thus proceeds between the nucleus and a target electron or a virtual photon similarly originating from the target. With this method the charge distribution may be measured, in this case of the projectile, which might be a nucleus of very short lifetime, $\gamma c\tau\simeq1$ mm, where $\gamma$ is the Lorentz factor, $c$ the speed of light and $\tau$ the lifetime.
With the proposal to extract protons and heavy ions from the LHC for fixed target physics, the so-called AFTER@LHC, such measurements would in principle enable charge distributions, or at least sphericity, for nuclei with lifetimes down to femtoseconds to be extracted. We report calculations of bremsstrahlung emission from Pb and Ar nuclei, with energies corresponding to the maximum of the LHC.

\section{Bremsstrahlung}
We study bremsstrahlung emission by relativistic heavy ions. 
When traversing an amorphous target, the projectile ions interact with the target electrons and nuclei. 
This causes radiation emission and energy loss to the projectiles. 
We focus on the radiation and assume the ion beam to be monoenergetic, that is, we consider targets sufficiently thin that the projectile energy loss is minimal.

To establish a reference value for the cross section, we first consider the incoming ion as a point particle of electric charge $Ze$ colliding with target atoms of nuclear charge $Z_te$.
The major part of the radiation is due to the interaction of the projectile with target nuclei which, in turn, are screened by target electrons at distances beyond the Thomas-Fermi distance, $a_{TF}$.
The cross section differential in energy for the emission of bremsstrahlung photons from an ion with atomic number $A$ then reads \cite{Jackson75}
\begin{equation}
\frac{d \chi}{d \hbar\omega} = \frac{16}{3}\frac{Z_t^2Z^4}{A^2}\alpha r_u ^2L,
\label{eq:Reference}
\end{equation}
where $\alpha \equiv e^2/\hbar c$ is the fine-structure, $e$ the unit electric charge and $\hbar$ the reduced Plancks constant.
The classical nucleon radius is defined as $r_u \equiv e^2/M_uc^2$, where $M_u$ is the atomic mass unit.
Expression (\ref{eq:Reference}) gives the radiation cross section or power spectrum; it is the number spectrum weighted by the photon energy, $\hbar\omega$. 
The factor $L$ is given by
\begin{align}
\begin{split}
 L  \approx \textrm{ln} \left( \frac{233M}{Z_t^{1/3}m} \right) - \frac{1}{2} \left[ \textrm{ln} \left(1+r^2 \right) - \frac{1}{1+r^{-2}} \right], \\
 r = \frac{96 \hbar \omega}{\gamma \gamma' Z_t^{1/3}mc^2},
\end{split}
\end{align}
where $\gamma \equiv E/Mc^2$, $\gamma' \equiv (E- \hbar\omega)/Mc^2$, $m$ is the electron mass and $M$ is the mass of the projectile. 
The material-dependent factor $L$ accounts for the electronic screening of the target nuclei.
It is essentially the logarithm of the ratio between the effective maximum ($\sim 2 Mc$) and minimum $\left( \sim \hbar/a_{TF} \right)$ momentum transfers  to the scattering center. 
The reference power spectrum extends all the way up to the energy of the primary ion and varies only slightly with energy.
However, as we shall study in this paper, photons with energy $\hbar\omega \gtrsim \gamma \hbar c/R$ have wavelengths smaller than the radius of the ions which cause the emission; making them sensitive to the nuclear structure and collective dynamics of the constituent protons.
Taking this into account causes significant change in the shape of the bremsstrahlung spectrum.

\section{The Weizs\"{a}cker-Williams method}
When the emitted bremsstrahlung photons have a wavelength that is small compared to the nucleus, the size and structure of the nucleus affect the emission.
We can investigate this by using the Weizs\"{a}cker-Williams method of virtual quanta \cite{Jackson75,Williams,Williams2,Weiz}. 
In this approach, we represent a moving charged particle by a spectrum of virtual photons which scatter on 
a stationary charged particle. There are contributions to bremsstrahlung from scattering both on target constituents and on the projectile ion. 
The latter is the dominant contribution.
Hence we change to a reference frame in which the incoming ion is at rest and where we represent the screened target nuclei by a bunch of virtual photons. 
These photons scatter off the ion and are subsequently Lorentz boosted back to the laboratory frame - resulting in an energy increase by a factor of up to $2\gamma$.
This means that the bremsstrahlung can be calculated using the Weizs\"{a}cker-Williams method of virtual quanta, photonuclear interaction theory and a Lorentz transformation.
For the cross section we take the elastic photonuclear interaction cross section as this ensures the scatterer to remain intact, that is, the incoming ion does not disintegrate in the process of radiation emission. 
The spectrum of virtual photons is given in \cite{Jackson75,Sore10}.
Multiplying this with the photonuclear scattering cross section differential in angle results in the scattering cross section differential in energy and angle.
The transformation to the laboratory frame is performed by utilizing an invariance relation \cite{Jackson75} and produces the bremsstrahlung power spectrum differential in energy and angle, for details see \cite{Sore10,Us15,JenSor2013}.

In Ref. \cite{Sore10}, one of us used this procedure to calculate the bremsstrahlung spectrum of relativistic bare lead ions.
This was possible by using the photonuclear interaction data provided in Ref. \cite{Schelhaas88}. 
However, data for other nuclei is not abundantly available.
We therefore developed a procedure to derive the necessary elastic scattering cross sections taking total photonuclear absorption cross sections as input; these are available in the ENDF database for about $100$ different nuclei \cite{ENDF}.
We obtain the elastic scattering cross sections at low to moderate energies, that is, at energies covering the giant dipole resonance by applying the optical theorem and a dispersion relation to the total photonuclear absortion data, see Ref. \cite{Us15}.
At higher energies additional constraints are invoked to ensure coherence.
With this construct, the bremsstrahlung-spectrum can be calculated for any ion for which the total photon absorption cross section is known.
Due to the available data, our approach is most exact for lead ions which have already been successfully accelerated to $4$ TeV$\times Z$ in the LHC machine. 
Also, this allowed us to cross check the earlier calculations for the bremsstrahlung from lead. 
It has not been finally decided if other ions will ever be used in the LHC.
But argon ions are frequently discussed as a possibility if the physics case requires lower mass ions to be accelerated \cite[vol.\ 3, ch.\ 33]{LHCdr} - see volume 3 chapter 33. 
Supporting this idea, in 2015, the CERN accelerator complex is successfully accelerating argon ions in the Super Proton Synchrotron at energies up to $150$ GeV/n.
We therefore also provide bremsstrahlung calculations for argon ions at LHC energy in the next section.

\section{Results - Bremsstrahlung}
\begin{figure}[hbt]
\begin{center}
\includegraphics[width=0.5\textwidth]{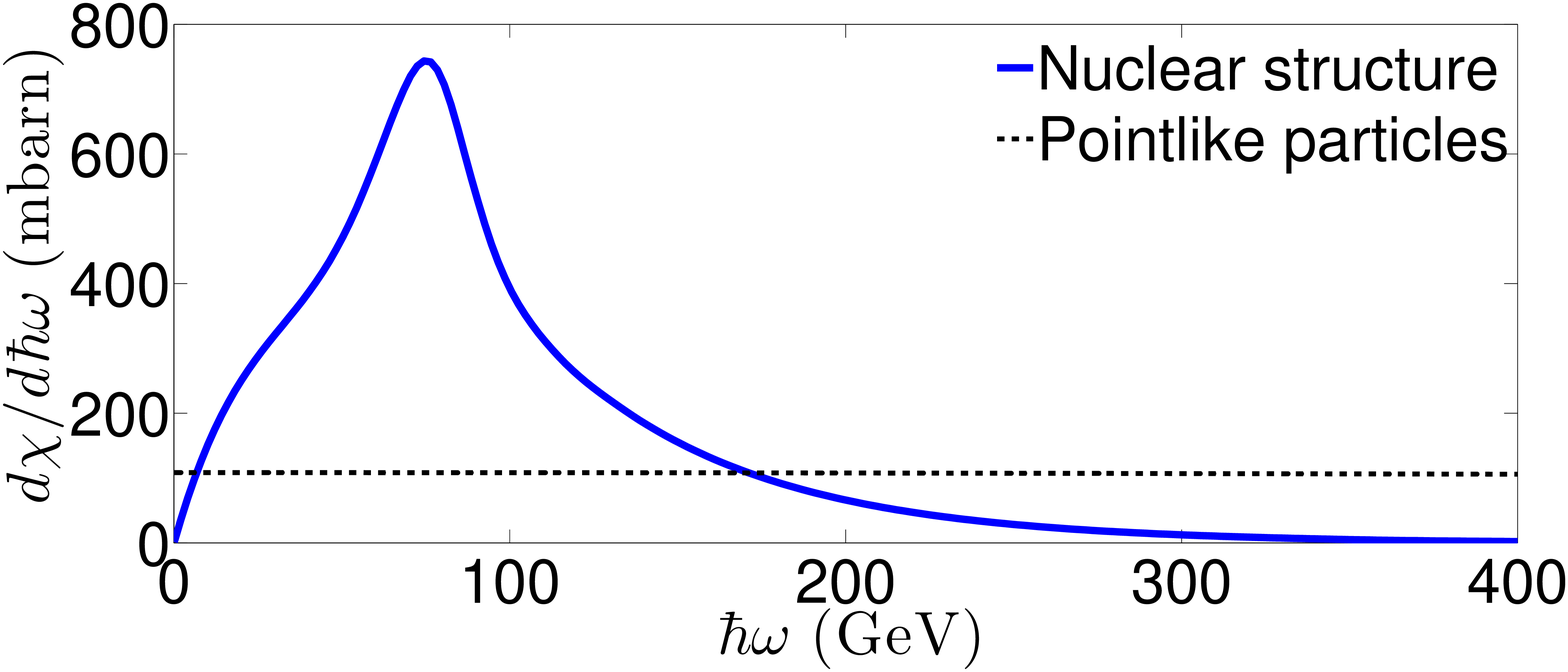}
\caption{Bremsstrahlung calculations for a Pb-208 projectile with energy $7$ TeV$\times Z$ incident on a lead target. The dashed line shows the reference cross section (\ref{eq:Reference}) and the full drawn curve shows the present results. }
\label{fig:leadGamma}
\end{center}
\end{figure}

In this section, we present calculations of bremsstrahlung spectra for bare argon and lead ions at $7$ TeV$\times Z$ incident on a fixed target.
Figure \ref{fig:leadGamma} shows the power spectrum of bremsstrahlung for lead ions aimed at a lead target. 
The spectrum obtained by integrating over all emission angles has a pronounced peak around a photon energy of about $80$ GeV. 
This corresponds to the collective interaction of the projectile nucleons with the virtual photons of the target - the giant dipole resonance.
This well-known resonance in the photonuclear cross sections is also apparent in the bremsstrahlung spectrum, albeit here multiplied by a factor of $2\gamma$ from the Lorentz boost. 
At energies above the peak, significantly fewer photons are predicted by the current model.
This is because coherence is restricted to a decreasing range of photon scattering angles such that most scattering events correspond to incoherent interaction with the projectile protons. 
Incoherent interaction of an energetic virtual photon with a target proton generally leads to breakup of the nucleus and hence does not contribute to the spectrum. 
This decreasing contribution from coherent scattering leads to a depletion of the elastic scattering cross section at higher energies, which is apparent also in the bremsstrahlung spectrum. 
Since the dashed line in Figure \ref{fig:leadGamma} extends all the way up to the energy of the ion, the integrated difference between the two curves is very large.

\begin{figure}[hbt]
\begin{center}
\includegraphics[width=0.5\textwidth]{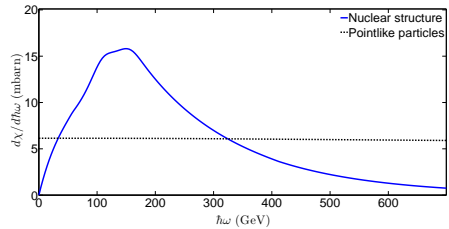}
\caption{Same as Figure \ref{fig:leadGamma} except for projectile which is here Ar-40.}
\label{fig:argonGamma}
\end{center}
\end{figure}

The spectrum for argon projectiles is shown on Figure \ref{fig:argonGamma}.
It is largely similar to that for lead although the overall values are of course much lower. 
The peak height is approximately $50$ times lower, and it must be noted that there is no simple scaling between the heights and shapes of the two spectra (scaling the heights with $Z^2$ off-shoots by a factor of $2-3$ here).
The lack of such scaling is traced back to differences in the photonuclear scattering cross sections.
The argon spectrum is somewhat broader than that for lead and the high energy tail extends to larger energies than for lead.
This difference is also present at the elastic scattering cross sections and is due to the different shapes of the argon and lead nuclei. 
Lead is almost spherically symmetric and this leads to a very narrow giant dipole resonance peak. 
For argon on the other hand, the nucleus is much less symmetric, and the photonuclear as well as the bremsstrahlung cross sections actually consist of two individual but closely lying peaks (for bremsstrahlung this will show through collimation). 

\begin{figure}[hbt]
\begin{center}
\includegraphics[width=0.5\textwidth]{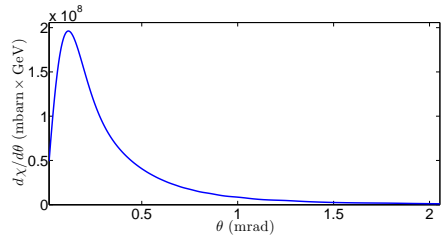}
\caption{The angular distribution of bremsstrahlung from a Pb-208 projectile with energy $7$ TeV$\times Z$ incident on a lead target. Note that the cross section is shown differential in the polar emission angle $\theta$ rather than in solid angle and hence includes a factor of $2\pi\textrm{sin}(\theta)$. The majority of the bremsstrahlung photons are indeed emitted in the very near forward direction.}
\label{fig:BSangularDistribution}
\end{center}
\end{figure}

Figure \ref{fig:BSangularDistribution} shows the bremsstrahlung spectrum obtained for lead ion by integration over emission energy, that is, differential in angle instead of energy. 
As expected, there is a peak in radiation intensity at an angle corresponding to $1/\gamma$. 
For LHC beam energies this means that the bremsstrahlung photons are emitted at angles of order less than $1$ mrad.

\section{Summary and Conclusions}
We have presented calculations on the bremsstrahlung emission from relativistic heavy ions with energy corresponding to that of the LHC beam. 
The calculations are novel in the way that knowledge on nuclear structure is taken into account using existing data on photonuclear cross sections. 
In addition, our approach has not before been applied to energies above that available at the SPS. 
We demonstrate that substantial cross sections for the emission of high energy photons are expected around energies of $\hbar\omega \approx$ $100$ GeV.
Here our model produces a radiation peak which overshoots the result for a pointlike particle of the same charge and mass by a factor of roughly 2 and 6 for argon and lead respectively. 
In the energy regions below and above the peak however, our model produces significantly fewer bremsstrahlung photons than what is otherwise expected. 
If this holds true against experiments, it implies that the energy loss of the ions through the bremsstrahlung channel is much less severe than previously expected by some authors, cf. \cite{Sore10}. 

Our calculations can of course also be performed at lower energies - see \cite{Us15} for calculated cross sections for energies of about 150 GeV/n.
The SPS can accelerate to about $450$ GeV$\times Z$ so that experiments located in the SPS fixed-target hall should be able to see this signal. 
The COMPASS experiment \cite{COMPASS} may be a candidate. 
See also \cite{ICPEAC} for calculations on delta electron emission from a similar experimental condition as discussed here. 
Along with bremsstrahlung measurements, such electrons would be highly sensitive to the nuclear charge structure.

If a fixed target facility using the LHC beams is constructed, one could hope for the option to produce secondary beams. 
With such beams, one could study the bremsstrahlung from short-lived nuclei.
If traversing a $1$ mm target, the lifetime would be $\gamma c \tau$ = $1$ mm $ \implies \tau \approx 4\cdot10^{-16}$ s. 

Exploiting the Lorentz time dilation, the nuclear structure of these rare nuclei could be exposed.  
Potentially, along with these short-lived species, one may also study exotic beams where the nucleus contains a strange quark.
It is presently unknown whether the presence of a strange quark increases or decreases the radius of the nuclear charge. 
Whereas such effects may be visible in the bremsstrahlung spectrum, this would certainly be impossible to study using conventional electron scattering.

\end{document}